\documentclass[a4paper,11pt]{article}
\pdfoutput=1 

\usepackage{jheppub} 

\usepackage[T1]{fontenc} 
\usepackage{amsmath,amssymb}

\usepackage{comment}

\newcommand{\ba}{\begin{eqnarray}}
\newcommand{\ea}{\end{eqnarray}}
\newcommand{\nn}{\nonumber}

\newcommand{\cO}{{\mathcal{O}}}

\newcommand{\be}{\begin{equation}}
\newcommand{\ee}{\end{equation}}

\def\a{\alpha}
\def\b{\beta}

\def\d{\delta}
\def\e{\epsilon}

\def\n{\nu}

\def\t{\tau}

\def\D{\Delta}

\def\L{\Lambda}

\allowdisplaybreaks[1]
\title{\boldmath Classical Virasoro irregular conformal block}

 \author{Chaiho Rim}
 \author{and Hong Zhang}
 \affiliation{Department of Physics and Center for Quantum Spacetime (CQUeST),\\
Sogang University, Seoul 121-742, Korea}

\emailAdd{rimpine@sogang.ac.kr}
\emailAdd{kilar@sogang.ac.kr}

\abstract{Virasoro irregular conformal block with arbitrary rank is obtained
for the classical limit or equivalently Nekrasov-Shatashvili limit
using  the beta-deformed irregular matrix model 
(Penner-type matrix model for the irregular conformal block).
The same result is derived using the generalized  Mathieu equation 
which is equivalent to the loop equation of the irregular matrix model. }

\begin{document} 
\maketitle
\flushbottom

\setcounter{footnote}{0}


\section{Introduction} 

Virasoro irrregular module, so called Gaiotto state or Whittaker state \cite{Whittaker}
turns out to be  in connection 
with the four dimensional $\cal N$=2 super Yang-Mills theory \cite{G_2009}. 
The irregular module is different from the regular one in that 
it is the simultaneous eigenstate of positive Virasoro generators $L_n$ with $n>0$ 
instead of $L_0$. 
According to the AGT conjecture \cite{AGT} 
there is a duality between 4D gauge theory and 2D conformal field theory: 
the instanton  partition sector of the Nekrasov partition of class $T_{g,n}$ 
in  $SU(2)$ quiver gauge theory is given 
as the Liouville conformal block on a Riemann surface with genus $g$ and $n$-regular punctures.
However, there can arise irregular punctures whose degree is higher than 2 
when the number of flavors is less than the regular one.
In this case, the partition is constructed in terms of the inner product of Gaiotto states 
and provides the information on the Argyres-Douglas theory \cite{{AD},{APSW}}. 

On the other hand, it is noted that 
the irregular conformal block (ICB) is obtained from the colliding limit 
of the regular conformal block \cite{{EM_2009}, GT_2012}. 
The colliding limit is the fusion of primary vertex operators 
when the Liouville charge is very big
so that their moment constructed as the product of the charge with 
powers of its position is finite. 
The ICB  can be easily studied in terms of Penner-type matrix model 
which consists of logarithmic potential 
and finite Laurent series (finite number of positive powers and negative powers of matrix) \cite{CRZ}. 
We will call this matrix model hereafter, the irregular matrix model (IMM). 
This model can describe the inner product between Gaiotto states of arbitrary rank
and also provide the ICB. 

The Liouville conformal block can be also studied using the classical limit.
The 3-point conformal block was studied in \cite{ZZ} and 
the 4-point one is given in terms of  Painlev\'e VI in \cite{LLNZ_201309}.
From the AGT-conjecture point of view, 
the classical limit is equivalent to the Nekrasov and Shatashvili (NS) limit:
The $\e$ parameter of the Nekrasov partition function 
with the $\Omega$ background is identified with the Planck constant.
Nekrasov and Shatashvili obtained the lowest energy using the first quantization 
of  integrable systems  \cite{NS}.  
In addition, the classical limit was equivalently 
investigated in terms of $\beta$-deformed 
 Penner-type matrix model in \cite{BMT_201104}.

Considering the current trend of research
one may wonder if there is the classical limit of ICB. 
The simplest case  is obtained 
using the Bohr-Sommerfeld periods 
of one-dimensional sine-Gordon model \cite{MM_0910}.
Similarly, degenerate conformal block is used to 
obtain the information on the classical limit of the ICB 
in \cite{LLNZ_201309, PP_201407}
In this paper, we will provide NS limit 
of the inner product of irregular modules 
for arbitrary rank 
using the IMM and evaluate the partition function in a systematic way.

This paper is organized as follows. 
In section 2,  the NS limit of the IMM  is provided.
We review the property of the beta-deformed regular Penner-type matrix model
(regular matrix model) 
and obtain the NS limit. The NS limit of the IMM is 
obtained from the colliding limit of the NS limit of the regular 
matrix model. It is noted that the order of the NS limit and the colliding limit is
immaterial. One may equally take the colliding limit first and the NS limit later. 
We present the NS limit first in this section 
since NS limit of the regular conformal block appeared 
in  \cite{BMT_201104}  already. 
In section 3, the NS limit of the IMM is identified with the NS limit of the 
irregular module with arbitrary rank
and the explicit representation is provided.
In addition, the exponentiated form  is provided 
for the NS limit of  ICB . 
Furthermore,  explicit form of the dominant contribution 
 is provided for the arbitrary rank case. 
In section 4, the same result is obtained with a slightly different method.
We use the degenerated primary operator method to find a second order
differential equation for the rank 1. 
As a result, the Schr\"odinger equation is obtained 
which has two cosine potentials. 
We find a similar result for the arbitrary rank whose derivation 
is given using the IMM. 
A differential equation is obtained 
with the combination of exponential (cosine) terms 
which will be called the generalized Mathieu equation. 
We provide a check of the result obtained in section 3
with that from the second order differential equation. 
Section 5 is the summary and discussion. 

\section{Irregular  matrix model and NS limit } 

\subsection{Liouville conformal block and regular matrix model} 
Liouville conformal block  is given as the holomorphic part of the 
correlation with  $N$  number of screening operators  
\be
{\cal F}^{(n+2)}=\left \langle\!\!\!\left \langle  \left( \int d \lambda_I e^{2b \phi(\lambda_I)} \right)^{\!\!N}  ~~
         \left(  \prod_{a=0}^{n+1} e^{2 \a_a \phi(z_a)} \right) \right \rangle \!\! \!\right \rangle\,,
\ee
where the double bracket stands for the expectation value with respect to free field action.
Using the free field correlation  
$\left \langle\!\!\!\left \langle \phi(z) \phi(\omega) \right \rangle \!\! \!\right \rangle
 \sim - \frac{1}{2} \log (z - \omega)$
one has the conformal block in Selberg integrals:
 \be 
{\cal F}^{(n+2)}  =  \prod_{0\leq k<I\leq n+1} (z_k - z_l)^{-2 \a_k  \a_l}  ~
\int  \left (  \prod_{I=1}^N  d \lambda_I \right )
  \Delta(\{\lambda_I\})^{-2b^2}   \prod_a  (\lambda_I - z_a)^{-2b \a_a} \,,
 \ee
where 
$ \Delta(\{\lambda_I\})= \prod_{I < J} (\lambda_I - \lambda_J) $ is the Vandermonde determinant.

We will fix the position  $z_0=0$ as the reference point 
and put $z_{n+1} \to \infty$. 
The $\beta$-deformed Penner-type matrix model \cite{DV_200909, {Itoyama}} is defined 
from the conformal block by factoring out 
the prefactor $  \prod_{0\leq k<I\leq n+1} (z_k - z_l)^{-2 \a_k  \a_l}$ 
as well as the terms containing $z_{n+1}$:
\be
Z_{\beta}   =     \int \left( \prod_{I=1}^N d \lambda_I \right)
 \prod_{I < J} (\lambda_I - \lambda_J)^{2\b} 
            e^{ \frac{\sqrt{\beta}}{g}  \sum_I V(\lambda_I)}\,,
\label{eq_beta-matrix}
 \ee
with $b= i \sqrt{\beta}$ and the Penner potential $V(\lambda)$ 
is defined as 
\be
  V(\lambda )  =     \sum_{a=0}^{n} \hat \a_a \log (\lambda - z_a)\,.
\label{eq_potential}
\ee 
Here we introduce the scaling parameter $g_s$ so that  
$\alpha$ is scaled:
$ \a_k \equiv \hat \a_k / g_s$ for later convenience.
The expansion parameter $g$ is related with $g_s$ as $g = ig_s/2$ 
so that $\sqrt{\beta} /g = - 2 b /g_s$. 
The $\beta$-deformed matrix model  reduces to the hermitian  matrix model 
when $\beta=1$.
We call this beta-deformed Penner-type matrix model the regular matrix model
to distinguish from the	IMM. 

The Liouville charge $\hat \alpha$ satisfies the neutrality condition 
 \be
   \sum_{a=0}^{n+1} \hat \a_a    + b g_s N = g_s Q\,,
\label{eq_neutrality} 
 \ee
where $Q$ is the Liouville background charge 
$    Q=b+    {1}/ {b}  $ and $bg_s = 2g \sqrt{\beta} $. 
Be aware that 
the Liouville charge $ \hat \a_{n+1}$ located  at $\infty$  
is included in the neutrality condition even though the matrix model does not contain
the term at  the spacial infinity. 
If one takes the conformal transformation $\lambda \to 1/\lambda$, 
then the Liouville charge at infinity 
appears naturally through the neutrality condition. 

It is well-known that the symmetric property of the matrix model 
is given in terms of the loop equation which corresponds to the Ward-identity \cite{TR_201207}: 
\be
4 W(z)^2 + 4 V'(z) W(z) + 2  g_s Q W'(z) -  g_s^2 W(z,z) =f(z) \,,
\label{eq_loop}
\ee
where $W(z_1, \cdots, z_s)$ is the $s$-point resolvent and 
is defined as 
\be
W(z_1, \dots, z_s)= \beta \left( \frac{g}{\sqrt{\beta}} \right)^{2-s} 
\left\langle 
\sum_{I_1, \cdots, I_s } \frac1 {(z_1-\lambda_{I_1})  \cdots (z_s-\lambda_{I_s})  }  
\right\rangle_{\!\! conn}\,.
\label{eq_multi-resolvent}
\ee
Explicitly one and two point resolvents are given as 
$ W(z) = g \sqrt{\beta}   \left\langle\sum_I  \frac{1} {z - \lambda_I} \right\rangle_{\!\!\!conn}$
and
$ W(z,w) = \beta   \left\langle\sum_I 
 \frac{1} {(z - \lambda_I)(w - \lambda_I) } \right\rangle_{\!\!\!conn}$
respectively. 
The bracket  $\left\langle O \cdots \right \rangle_{\!\!\!conn} $denotes the  
connected part of the  expectation value of  $ O \cdots $ with respect to the matrix model 
\eqref{eq_beta-matrix}. 

$f(z)$ is the  expectation value determined by  the potential $V(z)$ 
\be
 f(z)    =   4g \sqrt{\beta} \left\langle
 \sum_I \frac{V'(z)- V'(\lambda_I)}{z - \lambda_I} 
\right\rangle_{\!\!\!conn}\,.
\label{eq_f}
\ee 
With the Penner type potential in  \eqref{eq_potential},  
$   f(z)    =     \sum_{a=0}^{n} \frac{d_a}{z - z_a}$  
and $d_a$ is determined by the derivatives of the partition function 
\be
    d_a
     =   - 2 b g_s \left< \sum_I \frac{ \hat \a_a}{\lambda_I - z_a} \right>
     =  -   g_s^2 \frac{\partial \log Z_\beta }{\partial z_a}\,.
\label{eq_da}
\ee
One may rewrite the the loop equation \eqref{eq_loop} as 
\be
x(z)^2 + g_s Q x'(z) -g_s^2 W(z,z) = -g_s^2 \varphi(z) \,,
\label{eq_loop-x}
\ee 
where
$x(z) = 2 W(z) + V'(z)$ and  $\varphi(z)$ turns out to be the expectation value 
of the energy-momentum tensor whose explicit form is given as \cite{TR_201207}
\begin{align}
\varphi(z)
& \equiv
 -\frac1 {g_s^2}\bigg\{( V'(z))^2 + g_s Q V''(z) + f(z) \bigg\}
\nn\\
&=\sum_a  
\left( 
\frac{\Delta_a} {(z-z_a)^2 } + \frac1{z -z_a} \frac{\partial \log Z_{eff}} {\partial z_a} 
\right)\,,
\end{align}
with $Z_{eff}= Z_\beta\times \prod_{0\le a <b\le n} (z_a -z_b)^{-2 \alpha_a \alpha_b}$.

The Liouville conformal block has a direct relation \cite{BMT_201104}
with gauge theory through the AGT conjecture.
One may put the position of the vertex operators
$z_0 = 0$, $z_1 = 1$ and
$   w_a  =      q_1 q_2 \ldots q_{a-1}$ for $a=2, \cdots, n$ 
and  $q_a =e^{2 \pi i \tau_a}$
with  $\tau_a$  the gauge coupling constants.
Then one has   $ d_a    =      ( u_{a-1} - u_{a} )/ (2 \pi i z_a)$
where  $u_{a} = -g_s^2   {\partial \log Z}/ {\partial \tau_a }$.
Here  the relation 
 $2 \pi i z_a \frac{\partial}{\partial w_a} 
= \frac{\partial}{\partial \tau_{a-1}} - \frac{\partial}{\partial \tau_a}$
is used.
In this way, $d_a$'s  relate with  the Higg's field  expectation value 
$\langle {\rm Tr}\, \Phi^2 \rangle$. 

\subsection{Regular matrix model and NS limit} 

The $\Omega$ deformation parameters 
$\epsilon_1$ and $\epsilon_2$
of the gauge theory  are related with  
the Liouville parameter $b$.
One may identify $   \epsilon_1      =     b $ and $   \epsilon_2     =   1/b$
so that  $\epsilon_1 \epsilon_2 =1$. 
On the other hand, one may define  the gauge theory  
in the NS limit
where  $\epsilon_2=0$ but $\epsilon_1$ is finite.
To get the corresponding limit for the matrix model
one may define the parameter relation in a different way so that 
the NS limit is obtained easily. 
To achieve this, one may rescale  $\Omega$ deformation parameters
so that one has 
 \be
    \epsilon_1  =   g_s b, ~~~    \epsilon_2     =     \frac{g_s}{b}\,,
 \ee
which provides the overall scale $ \epsilon_1 \epsilon_2 = g_s^2 $.  
In this new convention, NS limit ($\epsilon_2 \to 0$ and $\epsilon_1$ finite)
corresponds to the limit   $g_s \to 0$ and $b\to \infty$. 
Note that the Liouville theory has $b \to 1/b$ duality
and this duality for the gauge theory is $\epsilon_1 \to \epsilon_2$. 
Therefore, NS limit is equivalent to the classical limit $b \to 0$.
Or one may equally put $\epsilon_1 \to 0$ instead of $\epsilon_2$.
In this way, we will not distinguish the NS limit from the classical limit.
 
Under the NS limit, physical quantities are to be rescaled properly.
The background charge scales as  $Q \to   \epsilon_1 /g_s$ 
and the central charge $c=1 +6 Q^2\to 6\epsilon_1^2/g_s^2$.  
The conformal dimension $\Delta (\alpha) = \alpha(Q-\alpha)$
of primary operator  will scale as  
$\Delta (\alpha) =\delta_\alpha /g_s^2 $ 
where $\delta_\alpha= \hat \alpha( \epsilon_1 -\hat \alpha)$ 
with  $ \hat \alpha  $ finite. 
In addition, the neutrality condition \eqref{eq_neutrality}
maintained with no scaling of $N$. 

Note that the potential in   \eqref{eq_potential} 
dose not change but the power $\beta$ of the Vandermonde determinant 
should scale as $ \beta \to  -  \epsilon_1 b/g_s $
and  $g_s Q \to   \epsilon_1 $. 
Besides, in NS limit  the 2-point resolvent 
$g_s^2 W(z,z)$ vanishes since $W(z,z)$ is finite \cite{MMM}.
As a result, the loop equation \eqref{eq_loop-x} 
has the form   
\be
       x(z)^2 +  \epsilon_1 x'(z)  + U(z) =0 \,,
\label{eq_loop-i}  
\ee 
and $U(z)$ is the NS limit of $g_s^2 \varphi(z) $
\be 
 U(z) =     \sum_{a=0}^{n} \frac{\delta_{\alpha_a}}{(z - z_a)^2}
         + \sum_{a=0}^{n} \frac{\chi_a}{z - z_a}\,.
\ee 
The coefficients are defined as  
 $\delta_{\alpha_a}  = \hat \alpha_a ( \epsilon_1 -\hat \alpha_a)$
and $\chi_a =   \sum_{ b (\neq a)}  {2 \hat \a_a \hat \a_b}/{(z_a - z_b) }-d_a$
where $d_a$ is defined in \eqref{eq_da}. 

It is noted that the loop equation turns into the second order 
differential equation with $n$ regular singularities present in $U(z)$:
    \be
   \left ( \epsilon_1^2 \frac{\partial^2}{\partial z^2}   + U(z) \right)  \Psi(z) =0\,,
\label{schrodinger}
    \ee 
where  $   \Psi (z)
     =     \exp \left( \frac{1}{\epsilon_1} \int^{z} x(z') dz' \right)$.
Therefore, one may view the loop equation \eqref{eq_loop-i}  
as the Hamilton-Jacobi like equation.
In \cite{LLNZ_201309}, the 4-point classical block ($n=4$) 
with the position identified as ($0,1, t, \infty$)
is converted into the conventional Hamilton-Jacobi equation 
and the function $z(t)$ is noted to satisfy 
the Painlev\'e VI.

\subsection{Colliding limit and irregular matrix model} 
\label{sec_ns-collid}

The colliding limit is used to find the ICB,
where many primary vertex operators are put at 
the reference point ($\,z_k\to z_0=0$) but with the Liouville charge infinite 
( $\hat \alpha_k\to \infty$)  so that their products have finite results, 
$ \hat c_k=\sum_{a=0}^n    \hat \a_a (z_a)^k, (k =0, 1, \cdots, n) $. 
The colliding limit of $n+1$ vertex operators
provides the maximum number  $n+1$ 
of the non-vanishing moments  $ c_k$
and  the potential  of the form
\be
 V (z ;  \{\hat c_k\}) = \hat  c_0 \log z
- \sum_{k=1}^{n} \frac{  \hat c_k}{k z^k}\,.
\label{eq_potential-i-n}
\ee 
$f(z)$ is defined in \eqref{eq_f} and  has the form \cite{TR_201207},
\be
f(z) =
-g_s^2 \sum_{k=0}^{n-1} \frac {v_k ( \log   {Z})} {z^{2+k}} \,,
\ee 
where  $ v_k =  \sum_\ell  \ell~\hat c_{\ell+k }  \frac{ \partial}{\partial \hat c_\ell} $.
Here  the notation  $\hat c_\ell=0$ is used when $\ell \ge n+1$. 
One has no term proportional to $1/z$  
 due to the identity $\langle \sum_I  \hat V'(\lambda_I) \rangle =0$.
The loop equation \eqref{eq_loop-i} maintains the same form 
and  $  U(z) $ is given explicitly as 
 \be
\label{uz}
    U(z)
     =    \sum_{k=0}^{2n} \frac {\Lambda_k}{z^{k+2}}
+ \sum_{k=0}^{n-1} \frac {v_k ( g_s^2 \log  \hat {Z})} {z^{2+k}} \,,
\ee
where 
 $\Lambda_k 
= (k+1) \epsilon_1 \hat  c_k-\sum _{\ell=0}^{n} \hat c_\ell \hat c_{k-\ell}$. 
Noting that $v_k$ is the representation of Virasoro operators 
in $\{c_k\}$ space \cite{GT_2012}, one realizes that 
$U(z)$ is the expectation value of the energy momentum tensor  
$T(z)$ where non-negative moment is non-vanishing. 
Therefore, one  can define the non-negative moment of Virasoro generators 
as $ {\cal L}_k = \Lambda_k + v_k$ with $k=0, 1, \cdots, 2n$
with the notation $v_k =0 $ if  $k=n, \cdots, 2n$. 
This identification realizes the Virasoro commutation relation  
on the irregular module 
\be 
[{\cal L}_k,  {\cal L}_\ell] = (\ell-k) {\cal L}_{k+\ell}\,.
\ee 
The new feature is that the Virasoro generator has 
non-vanishing expectation values $\Lambda_k$ when  $k=n, n+1, \cdots, 2n$.
This demonstrate that IMM is based on the rank $n$ 
 irregular module  which is defines as 
\be 
\label{coherent}
{\cal L}_k |I_n\rangle = \Lambda_k|I_n\rangle \,,
\ee 
 when  $k=n, n+1, \cdots, 2n$.
The partition function is identified as 
the NS limit of the inner product between a regular module located at infinity 
and a rank $n$ irregular module located at origin.

\section{Classical irregular conformal block} 
\subsection{Irregular conformal block and NS limit}

The partition function given  in  section \ref{sec_ns-collid} 
is the NS limit of the inner product between a regular module located at infinity 
and a rank $n$ irregular module located at origin. 
On the other hand, ICB is given as the inner product 
between two different irregular modules, one at the origin and one at infinity. 
It is noted that  NS limit of IMM
is equivalent to the colliding limit of NS limit of regular matrix model:
The order of the two limiting procedure is commutative.
One can equally take the colliding limit first and NS limit next. 
In this section we present the classical conformal block  
by taking the NS limit to the IMM. 

The partition function corresponding to the inner product between 
irregular modules has the potential of the  form   
\be 
\frac1 {g_s} V_{(m:n)}   (z; c_0, \{c_k\} , \{c_{-\ell}\})
=   c_0 \log z - \sum_{k=1 }^{n}   \left( \frac {{ c_k} } {k z^{k}} \right)   
+ \sum_{\ell=1}^{m}   \left( \frac { c_{-\ell}~ z^{\ell}}   {\ell}  \right )  \,,
\ee
where  $  c_0= \sum_{r=0}^{n} \a_r $ and $  c_k= \sum_{r=1}^{n} \a_r(z_r)^k  $ are the moments. Positive $k$ corresponds to the contribution at the origin and 
 negative $k$ at infinity. 
The ICB $
{\cal F}_\Delta^{(m:n)}  (\{c_{-\ell }  :  c_k \})$ 
 is obtained from the normalized inner product 
between irregular module so that 
$ 
{\cal F}_\Delta^{(m:n)}  (\{c_{-\ell }  :  c_k \})
= \langle I_m | I_n \rangle / (  \langle I_m | \Delta \rangle \langle \Delta|  I_n \rangle) $ 
where $|\Delta \rangle$ is the regular module with conformal dimension $\Delta$. 
The explicit form of the ICB is given using  the IMM  \cite{CRZ}: 
\be
{\cal F}_\Delta^{(m:n)}  (\{c_{-\ell }  :  c_k \})= 
\frac{e^{\zeta_{(m:n)}}   Z_{(m:n)}  (c_0; \{c_k\}; \{c_{-\ell}\})} 
{Z_{(0:n)}(c_0;  \{ c_k \})  Z_{(0:m)} ( c_\infty; \{ c_{-\ell }\}) }\,,
\ee 
where extra factor $e^{\zeta_{(m:n)}}$ is needed 
due to  the limiting procedure $z_a \to \infty$ and $z_b \to 0$. 
Note that original conformal block has  the factors  
$\prod_{a,b}  (z_a -z_b)^{-{2 \alpha_a \alpha_b}}  $ 
which we factored out but the limiting procedure 
 results in the finite contribution, so called $U(1)$ contribution 
 $  e^{\zeta_{(m:n)}} $, 
where $\zeta_{(m:n)} =\sum_k^{{\rm min}(m,n)} 2 c_k c_{-k}/k$.
Therefore, to have the right conformal block
we need to include this extra factor in the definition. 
In addition,  $ \langle I_m| \Delta \rangle $ is expressed 
 as $ Z_{(0:m)} ( c_\infty; \{c_{-\ell }\})$ 
with the change of variable $\lambda_i \to 1/\lambda_i$.

The evaluation of ICB is done in \cite{CR}. Note that 
the information of the irregular module at the origin is obtained 
if one regards  the potential  $V_0 = V _{(0:n)} (\{\lambda_i \} ; c_0, \{c_k\})  $ 
as the reference one and $\Delta V _0$ as its perturbation:
\be 
 \frac1 {g_s}V_0 
=\sum_{I=1} ^{N_0} \Big( 
c_0 \log \lambda_I -  \sum_{k=1}^{n} \frac{c_k} {k } \lambda_I^{-k} \Big)
\,;~~~ 
\frac1 {g_s} \Delta V _0  =\sum_{I=1} ^{N_0} \Big( 
  \sum_{\ell=1}^{n} \frac { c_{-\ell}}{\ell}  { \lambda_I^{\ell}}  \Big) \,.
\ee
That is, $V_0 $ is the potential for the partition function $Z_{(0:n)} $ with 
$N_0 (\le N)$  number of screening operators. 
At infinity one has  the reference potential 
$ \sum_{J=1} ^{N_\infty} \Big( 
c_0 \log \lambda_J  - \sum_{\ell=1}^{n} { c_{-\ell}}{ \lambda_i^{\ell}} /\ell  \Big)$ 
and its perturbation $  -\sum_{J=1} ^{N_\infty }  
\Big( \sum_{k=1}^{n} {c_k}  \lambda_J^{-k}/k \Big) $.
We introduce the number  $N_\infty $ of screening operators  at infinity 
so that  $N_\infty  +N_0 =N$. 

The ICB is obtained using the perturbation theory. 
For example, ICB for rank 1 is given  in power of $\eta_0 \equiv c_1 c_{-1}$  
 up to order  $ \cO (\eta_0^2 )$ as 
\be 
\label{f11}
{\cal F}_\Delta ^{(1:1)}=1+\eta_0  \frac{2\bar{c_0} \bar{c_\infty}}{\D} 
+\eta_0^2\frac{ {4\bar{c_0}^2\bar{c_\infty}^2 c}/{\D} 
+4\D+2+12(\bar{c_0}^2+\bar{c_\infty}^2)
+32\bar{c_0}^2\bar{c_\infty}^2}{c+2 c \D+2 \D(8 \D-5)}\,,
\ee
where $\bar c_k = Q- c_k$ . 
This is compared with the Gaiotto notation 
\be
\langle \widetilde{G_2}|\widetilde{G_2} \rangle=
1+\L \L' \frac{m m'}{2\D}
+ (\L \L')^2
\frac{  {m^2 m'^2 c}/{4\D} +4\D+2-3 (m^2+m'^2)+2 m^2 m'^2 }{c+2 c \D+2 \D(8 \D-5)}\,,
\ee
and gets the parameter relation 
$\Lambda^2= - c_1^2$ and $m \Lambda=2 c_1 \bar{c_0}$,

To have the  NS limit, one has the scaling 
 $ c_k = \hat c_k  /g_s$ and
 $\eta_0=\hat\eta_0  /g_s^2 $. 
Following the conjecture in  \cite{ZZ}, one may put 
the classical ICB in an exponentiated form 
\begin{equation}
{\cal F}_\Delta^{(1:1)}  \left(\eta_0\right)
\;\stackrel{g_s \to 0}{\sim}\;
\exp\left\lbrace \frac{1}{g_s^2} f_{\delta}\left(\hat\eta_0\right)\right\rbrace.
\label{eq_e-conjecture}
\end{equation}
If one expresses the exponentiated term $f_{\delta}\left(\hat\eta_0\right)$ in the 
power series in $\hat\eta_0 $, one has 
\be 
f_{\delta}\left(\hat\eta_0\right)
= 
\lim\limits_{g_s \to 0} g_s^2 \log {\cal F}_\Delta^{(1:1)}  \left(\eta_0\right)
=
\sum\limits_{n=1}\left(\hat\eta_0\right)^{n} f_{\delta}^{(n)} \,.
\ee
Explicit result is obtained from  \eqref{f11} 
\be 
f_{\delta}^{(1)} \;=\;  \frac{2\hat{\bar{c_0}} \hat{\bar{c_\infty}}}{\d} ,
~~~~ 
f_{\delta}^{(2)}\;=\; \frac{ {(10-\hat c){\hat{\bar{c_0}}}^2{\hat{\bar{c_\infty}}}^2 }
+2\d^3+6\d^2({\hat{\bar{c_0}}}^2+{\hat{\bar{c_\infty}}}^2)
}{( c+8 \d)\d^3}\,, 
\ee
or in  Gaiotto variables 
\be 
\label{fdelta}
f_{\delta}^{(1)} \;=\;  \frac{\hat{m} \hat{m'}}{2\d} ,
~~~~ 
f_{\delta}^{(2)} \;=\; \frac{ {(5 \d -3{\e_1^2})\hat{m}^2 {\hat{m'}}^2 }
-12(\hat{m}^2+ {\hat{m'}}^2)\d^2+16\d^3
}{16\d^3( 4 \d +3{\e_1^2})}\,.
\ee
\subsection{Dominant behavior of the exponentiated term}  
The conjecture \eqref{eq_e-conjecture}  equally  applies to 
higher ranks.  
We explicitly check this conjecture for  the dominant part at the NS limit. 
Suppose   ICB   at the NS  limit is 
exponentiated. Then one expands the ICB in powers of 
expansion parameters. 
For example, rank 1  has the expansion parameter  $\hat\eta_0$ and 
has the form 
\ba
&&{\cal F}_\Delta^{(1:1)}  \left(\eta_0\right)
\;\stackrel{g_s\to 0}{\sim}\;
\exp\left\lbrace\frac{1}{g_s^2}\sum\limits_{n=1}\left(\hat\eta_0\right)^{n} f_{\delta}^{(n)} \right\rbrace \nn \\
&&=1+\hat\eta_0\frac{1}{g_s^2} f_{\delta}^{(1)}
+\left(\hat\eta_0\right)^{2}\left\{ \frac{1}{g_s^2} f_{\delta}^{(2)} +\frac{1}{2}\frac{1}{g_s^4}\left(f_{\delta}^{(1)}\right)^{2} \right \}\nn \\
&&+\left(\hat\eta_0\right)^{3}\left\{ \frac{1}{g_s^2} f_{\delta}^{(3)} +\frac{1}{g_s^4}f_{\delta}^1f_{\delta}^{(2)} +\frac{1}{3!}\frac{1}{g_s^6}\left(f_{\delta}^{(1)}\right)^{3}\right \}+\dots
\ea
and can be compared with the ICB and finds the NS limit 
by putting 
{\small
\ba
{\cal F}_\Delta^{(1:1)}  \left(\eta_0\right)
=\sum\limits_{n=0}\left(\eta_0\right)^{n} F_{\Delta}^{(n)}(g_s^2) =
1+\hat\eta_0\frac{1}{g_s^2} F_{\Delta}^{(1)}
+\left(\hat\eta_0\right)^{2}\frac{1}{g_s^4} F_{\Delta}^{(2)}+\left(\hat\eta_0\right)^{3}\frac{1}{g_s^6} F_{\Delta}^{(3)}+\dots
\ea}
The dominant contribution is given as   $\lim\limits_{g_s\to 0} F_{\Delta}^{(n)} =\frac{1}{n!} \left(f_{\delta}^{(1)}\right)^{n} $. 
Indeed, the first few terms show the exponentiated behavior:
$ \lim\limits_{g_s\to 0} F_{\Delta}^{(1)} 
 =  {2\hat{\bar{c_0}} \hat{\bar{c_\infty}}}/ {\d}
= f_{\delta}^{(1)}  $
 and $\lim\limits_{g_s\to 0} F_{\Delta}^{(2)} =
  {2(\hat{\bar{c_0}} \hat{\bar{c_\infty}})^2}/{\d^2} 
=  (f_{\delta}^{(1)} )^{2}/2$. 

One may find more rigorous proof for this dominant behavior.
Let us first consider the simplest case (the rank 1/2) given in \cite{PP_201407}.
(This corresponds to  $N_f=0$, $ SU(2)$ and is  obtained using some appropriate limit from the rank 1 case.)
\begin{eqnarray}
{\cal F}_{c,\Delta}(\Lambda) 
 =\sum\limits_{n=0}\Lambda^{4n} \left[G^{n}_{c,\Delta}\right]^{(1^n) (1^n)}
 =\sum\limits_{n=0}\left(\frac{\hat\Lambda^4}{g_s^2}\right)^{n} \frac{1}{g_s^{2n}} \left[G^{n}_{c,\Delta}\right]^{(1^n) (1^n)}\,.
\end{eqnarray}

Setting the scaling $\Lambda=\hat\Lambda/g_s $ one has 
to demonstrate the dominant behavior
\ba
\lim\limits_{g_s\to 0} \frac{1}{g_s^{2 n}}\left[G^{n}_{c,\Delta}\right]^{(1^n) (1^n)} =\frac{1}{n!} \bigg(\lim\limits_{g_s\to 0}\frac{1}{g_s^2}\left[G^{n}_{c,\Delta}\right]^{(1) (1)} \bigg)^{n}.
\ea 
We use the property of the Gram matrix. Gram matrix is defined at each  level 
as follows:
\\
\medskip\noindent
At level $1$ : $\lbrace L_{-1}|\,\nu_\Delta\,\rangle \rbrace$,
\[
G^{n=1}_{c,\Delta} = \langle L_{-1}\nu_\Delta\,|\,L_{-1}\nu_\Delta\rangle
=\langle\nu_\Delta\,|\,L_{1}L_{-1}\nu_\Delta\rangle = 2\Delta.
\]

\bigskip\noindent
At level $2$: $\lbrace L_{-2}|\,\nu_\Delta\,\rangle,L_{-1} L_{-1}|\,\nu_\Delta\,\rangle\rbrace$,
\begin{equation*}
G^{n=2}_{c,\Delta} =
\begin{pmatrix}
\langle L_{-2}\nu_\Delta\,|\,L_{-2}\nu_\Delta\rangle  &  \langle L_{-1}^2\nu_\Delta\,|\,L_{-2}\nu_\Delta\rangle \\
\langle L_{-2}\nu_\Delta\,|\,L_{-1}^{2}\nu_\Delta\rangle & \langle L_{-1}^{2}\nu_\Delta\,|\,L_{-1}^{2}\nu_\Delta\rangle  \\
\end{pmatrix}
=
\begin{pmatrix}
    \frac{c}{2}+4\Delta  & 6\Delta \\
    6\Delta & 4\Delta(2\Delta+1) \\
\end{pmatrix}.
\end{equation*}

\bigskip\noindent
At level $3$: $\lbrace L_{-3}|\,\nu_\Delta\,\rangle,
L_{-2}L_{-1}|\,\nu_\Delta\,\rangle, L_{-1}L_{-1}L_{-1}|\,\nu_\Delta\,\rangle\rbrace$,
\begin{equation*}
G^{n=3}_{c,\Delta}=
  \begin{pmatrix}
    2c+6\Delta  & 10\Delta & 24\Delta \\
    10\Delta & \Delta(c+8\Delta+8) & 12\Delta(3\Delta+1) \\
    24\Delta & 12\Delta(3\Delta+1) & 24\Delta(\Delta+1)(2\Delta+1) \\
  \end{pmatrix},
\end{equation*}
and so on. 

At the NS limit,   the scaling behavior shows that  
$c=\hat c/ {g_s^2}$ and $\Delta= \delta / {g_s^2}$ are  of  the same order.
In addition, every commutator  $[L_{m},L_{-m} ]$ leads
 to the order of $ {\cal O}(\Delta)$.
Therefore, 
the term with  $\langle\nu_\Delta\,|\,L_{1}^nL_{-1}^n\nu_\Delta\rangle$
at level $n$ 
will result in  the  order of $ {\cal O}(\Delta)^n$, 
the highest order in the inverse powers of $g_s$
and provides the dominant behavior in the same row or column.  

Note that  $N \times N$ square matrix 
\begin{equation*}
A=
  \begin{pmatrix}
    a_{11}  & a_{12}  & \dots a_{1N}  \\
     a_{i1}  & a_{i2}  & \dots a_{iN}  \\
     a_{N1}  & a_{n2}  & \dots a_{NN}  \\
  \end{pmatrix},
\end{equation*}
has the cofactor matrix 
\begin{equation*}
 A^{\star}=
  \begin{pmatrix}
    A_{11}  & A_{21}  & \dots A_{N1}  \\
     A_{12}  & A_{22}  & \dots A_{N2}  \\
     A_{1n}  & A_{2n}  & \dots A_{NN}  \\
  \end{pmatrix},
\end{equation*}
where $A_{ij} $ is the cofactor of  $a_{ij} $ and 
the determinant of the matrix $A$ is given as  $ |A|=A A^{\star}$: 
$ a_{N1} A_{N1}+a_{N2} A_{N2}+\dots+a_{NN} A_{NN}=|A|$. 
Let's turn back to $G^{n}_{c,\Delta} $. In this case,
 \ba
\left[G^{n}_{c,\Delta}\right]^{(1^n) (1^n)}=\frac{A_{NN}}{|A|}.
\ea
As discussed in the above the NS limit picks up the $a_{NN} A_{NN}$ 
as the dominant  term and the Gram matrix reduces to 
\ba
\lim\limits_{g_s\to 0} \left[G^{n}_{c,\Delta}\right]^{(1^n) (1^n)} =\lim\limits_{g_s\to 0} \frac{1}{a_{nn} }
=\lim\limits_{g_s\to 0} \frac{1}{ \langle L_{-1}^{n}\nu_\Delta\,|\,L_{-1}^{n}\nu_\Delta\rangle }.
\ea 
Note that $ \langle L_{-1}^{n}\nu_\Delta\,|\,L_{-1}^{n}\nu_\Delta\rangle  
=  \langle \nu_\Delta\,|\,L_{1}^{n}L_{-1}^{n}\nu_\Delta\rangle$ is not 
simple to evaluate because one needs to apply the commutator repeatedly. 
For example, to calculate $\langle \nu_\Delta\,|\,L_{1}^{n}L_{-1}^{n}\nu_\Delta\rangle$, we need 
$[L_{1},L_{-1} ]=2L_{0}$, and $[L_{0},L_{-1} ]=L_{-1} $ an d so on. 
However at the NS limit the dominant term  simplifies since 
$[L_{0},L_{-1} ]=L_{-1} $ gives the lower contribution 
in  $ {\cal O}(\Delta)$. 
Therefore, at the NS limit, one can regard 
technically  $[L_{0},L_{-1} ]=0 $. 
In this case, one uses  the relation $[A ,B^n ]=n[A,B ]B^{n-1}$
where $A=L_1$ and $B= L_{-1}$. 
Then since  $[A,B ]$ commutes with $B$, one has 
$[L_{1},L_{-1}^n ]\,|\,\nu_\Delta\rangle=2n\Delta L_{-1}^{n-1}\,|\,\nu_\Delta\rangle$.
Repeatedly using this method, one concludes that 
\ba
\lim\limits_{g_s\to 0}  \langle L_{-1}^{n}\nu_\Delta\,|\,L_{-1}^{n}\nu_\Delta\rangle 
=\lim\limits_{g_s\to 0}( 2^n n! )\Delta^n\,.
\ea
As a  result,
\ba
\lim\limits_{g_s\to 0}\frac{1}{g_s^{2n}} \left[G^{n}_{c,\Delta}\right]^{(1^n) (1^n)} 
=\frac{1}{( 2^n n! )\delta^n}
=\frac{1}{n!} \bigg(\lim\limits_{g_s\to 0}\frac{1}{g_s^2}\left[G^{n}_{c,\Delta}\right]^{(1) (1)} \bigg)^{n}\,.
\ea 

To consider rank 1 case, remind that  the irregular module for rank $n$ is conjectured as \cite{BMT_2011,KMST_2013, CRZ} 
\be
| \widetilde {G_{2n}} \rangle =
\sum_{\ell,Y, \ell_p} 
\L^{\ell/n}
\left\{ 
\prod_{i=1}^{n-1}  a_i ^{\ell_{2n-i} } b_i^{\ell_i}  
\right\} 
m^{\ell_n} 
Q_{\Delta} ^{-1} \Big(
1^{\ell_1}2^{\ell_2}
\cdots 
 (2n-1) ^{\ell_{2n-1}} (2n) ^{\ell_{2n}}; Y \Big)
L_{-Y} |\Delta \rangle \,.
\label{G_2n}
\ee 
One has the ICB using this notation
\be
{\cal F}_\Delta^{(1:1)}
=\langle \widetilde{G_2}|\widetilde{G_2} \rangle
=\sum\limits_{k=0}(\L \L')^k
\sum\limits_{\ell_i, \ell_i'} {m'}^{\ell_1'} m^{\ell_1} 
Q_{\Delta} ^{-1} \Big(
1^{\ell_1}2^{\ell_2} ; 1^{\ell_1'}2^{\ell_2'} \Big)\,.
\ee
Using the scaling $\Lambda=\hat\Lambda/g_s $, 
$m=\hat m/g_s $, $\Delta=\delta/ {g_s^2}$, 
one has $ {\cal F}_\Delta^{(1:1)} = \sum\limits_{k=0}
\left( {\hat\L  \hat\L'/}{g_s^2}\right)^{k}F_2^{(k)} $ 
where 
\ba
F_2^{(k)}
 =
\sum\limits_{\ell_i, \ell_i'} ( {\hat m'}/{g_s})^{\ell_1'} (  {\hat m}/{g_s})^{\ell_1} 
Q_{\Delta} ^{-1} \Big(
1^{\ell_1}2^{\ell_2} ; 1^{\ell_1'}2^{\ell_2'} \Big) \,.
\ea 
The scaling   of the Gram matrix  results in the scaling of
$ F_2^{(k)}$ as $ {(\hat m \hat m')^k}/{( 2^k k! \delta^k)} $.
This concludes  $\lim\limits_{g_s\to 0} F_2^{(k)} 
=\frac{1}{k!} \left(\lim\limits_{g_s\to 0} F_2^{(1)}\right)^{k} $ 
which demonstrates  that the dominant  part of 
${\cal F}_\Delta^{(1:1)}  \left(\eta_0\right)$ 
is exponentiated. 

One may demonstrate the same behavior for arbitrary rank. 
However, for the rank $n \ge 2$, there appears one subtle behavior  
of the irregular module 
related with the coefficient $b_k$ in \eqref{G_2n}.  
In fact, $b_k$ is not a simple constant but is to be fixed  as \cite{CRZ}
$\Lambda^{k/n} b_k \to   \Lambda_k + v_k(\log  Z_{(0:n)} ) $. 
If one  understands $b_k$ as the replacement,
one can identify   $|\widetilde{G_{2n}} \rangle $  
with $ |I_n\rangle$ in  \eqref{uz}.
For notational convenience, we will  use either  
 $ |I_n\rangle$ or  $|\widetilde{G_{2n}} \rangle $ 
without distinction but its revised form is  tacitly assumed. 

The eigenvalues $\Lambda_k$ for $\vert I_n \rangle$ in 
are identified with the coefficients given in \eqref{G_2n}
by  the comparison between the two expression of irregular states $| \widetilde {G_{2n}} \rangle$ and $\vert I_n \rangle$ \cite{KMST_2013}:
\begin{align}
\Lambda_{2n-s}  
& = \frac{\langle \Delta| L_W L_{2n-s}| \widetilde {G_{2n}} \rangle}
 { \langle \Delta| L_W | \widetilde {G_{2n}} \rangle   }   
=\Lambda^{2n-s /n} a_s     
~~~~{\rm  for} ~0 \le s <n, 
\nn\\
 \Lambda_n 
&=\frac{
\langle \Delta| L_W L_{n}| \widetilde {G_{2n}} \rangle}
{ \langle \Delta| L_W | \widetilde {G_{2n}} \rangle}
=\Lambda m  \,,
\end{align} 
and 
$ 
\langle \Delta| L_W | \widetilde {G_{2n}} \rangle =\Lambda^{\ell /n}   
\left\{ \prod_{i=1}^{n-1}  a_i ^{\ell_{2n-i} } b_i^{\ell_i}  \right\} m^{\ell_n} 
$ 
when $W=1^{\ell_1} 2^{\ell_2} \cdots (2n)^{\ell_{2n}}$.

Putting  $\langle \widetilde{G_{2n}}|\widetilde{G_{2n}} \rangle 
 \equiv\sum\limits_{k=0}\left ({(\L \L')^{1/n} } /{g_s^2 } \right)^k   F_n^{(k)}$ 
one has
 {\small
\be
F_n^{(k)}
 ={g_s^2}^{k(n-1)/n}
\sum\limits_{\ell_i, \ell_i'} \left\{ 
\prod_{i=1}^{n-1}  a_i ^{\ell_{2n-i} } {a_i '}^{\ell_{2n-i} '}  b_i^{\ell_i} {b_i '}^{\ell_{i} '}
\right\} 
m^{\ell_n} {m'}^{\ell_n'} 
Q_{\Delta} ^{-1} \Big(
1^{\ell_1}
\cdots 
 (2n) ^{\ell_{2n}} ; 1^{\ell_1'}
\cdots 
 (2n) ^{\ell_{2n}'}\Big)\,.
\ee}
Since  the  parameters scale as  
$\Lambda=\hat\Lambda/g_s $, $m=\hat m/g_s $, 
$\Delta=\delta/ {g_s^2}$ and  
$ c_k= {\hat c_k}/{g_s}$
one has 
$a_i= {\hat a_i}/{(g_s^{i/n})}$, $b_i= {\hat b_i}/{(g_s^{2-i/n})}$
.
From this scaling one immediately finds 
the dominant contribution of the Gram matrix is
$Q_{\Delta} ^{-1} \Big( 1^n   ; 1^n \Big)$.
Therefore, one has 
$ F_n^{(k)} \to  {(\hat b_1 \hat b_1')^k }/  (k!   2^k \delta^k )$ 
which is the exponentiated form 
with 
$\lim\limits_{g_s\to 0} F_n^{(k)}= \left(\lim\limits_{g_s\to 0} F_n^{(1)}\right)^{k} \!\!\!/{k!}$.\\

\section{Classical irregular conformal block and second order differential equation} 
\subsection{Null vector approach}
 
In section 3  the NS limit of  ICB is shown  using IMM. 
In this section, we present  a different  approach to find the same quantity.
Conformal block with addition of a degenerate primary operator 
(degenerate conformal block)
satisfies the null condition, which is written 
as a differential equation. 
This method is used for the NS limit of the rank 1/2 
in \cite{LLNZ_201309, PP_201407}
to obtain the Mathieu equation.  

The degenerate primary operator $ V_{+}(z)\;\equiv\;V_{\Delta_+}(z)$ 
with the Liouville charge $\alpha = -1/(2b)$
has the conformal dimension 
$ \Delta_{+}\;=\;-\frac{1}{2}-\frac{3}{4b^2} $.
At level 2,  the null vector  arises:
\[
\chi_{+}(z)\;=\;\left[\widehat{L}_{-2}(z) -
\frac{3}{2(2\Delta_{+} +1)}
\,\widehat{L}_{-1}^{\,2}(z)\right]V_{+}(z)\,.
\]
The null vector needs to vanish when evaluated between any states, {\it i.e.}, 
$\langle\, I_{ \ell}\,|
\,\chi_{+}(z)\,|\, I_{k}\,\rangle\nonumber=0$. 
This provides the non-trivial constraint  
\be 
 \langle\, I_{\ell}\,|
\,\widehat{L}_{-2}(z)V_{+}(z)\,|\,I_{k}\,\rangle
+
{b^2}\,\langle\,I_{ \ell}\,|
\,\widehat{L}_{-1}^{2}(z)V_{+}(z)\,|\, I_{ k}\,\rangle\;=\;0.
\label{null}
\ee  

Let us consider the case of rank 1.  
Let us denote  degenerate irregular 3-point block as 
$
 \Phi(\Lambda, z)
=  \langle\, I_1;\Delta_L, m_L, \Lambda_L| V_{+}(z)| I_1;\Delta_R,m_R, \Lambda_R \,\rangle 
$
where each irregular module is assumed to be constructed 
with the highest state with conformal dimension $\Delta_L$ 
and   $\Delta_R$, respectively.
In addition, 
$m_{L,R}$ and $\Lambda_{L,R}$ are the eigenvalues characterizing the irregular module. 
However, we will restrict  ourselves to the case when all the $L$ parameters are 
the same with the $R$ parameters:
$\Delta_L=\Delta_R=\Delta$, $m_L=m_R=m$ and $\Lambda_L=\Lambda_R=\Lambda$.
In this case,  the constraint \eqref{null} reduces to  the second order differential equation,
\be 
\label{dphi}
\left[ 
b^2 \,z^2 \frac{\partial^2} {\partial z^2}
- \frac{3z}{2}\frac{\partial}{\partial z}
+\Lambda^2 \left(z^2+\frac1{z^2} \right)
+m \Lambda  \left( z + \frac{1} {z} \right)
+\frac{\Lambda}{2} \frac{\partial}{\partial\Lambda} 
+ \kappa
\right]
\Phi(\Lambda, z)  =0\,,
\ee
with $ \kappa =    \Delta - \Delta_+  /2 $. 
One may normalize the 3-point block 
considering the conformal dimension of the degenerate operator 
and inner product of the irregular model: 
\be 
 \Phi(\Lambda, z)\equiv
z^{-\Delta_+}   \langle\, I_1 |I_1 \rangle
  \psi (\Lambda, z)  \,.
\ee

Now we put  the inner product as  the exponential form
$  \langle\, I_1 |I_1\rangle \;\stackrel{g_s\to 0}{\sim}\;
\exp\left\lbrace\frac{1}{{g_s}^2}
f_{\delta}(\hat\Lambda)\right\rbrace $
and use the scaled quantities  
$\;\;\Delta_+ \to - {1}/{2}$ and $\Delta \to  \delta/ {g_s^2}$  
with $\delta\;=\;\e_1^2 (\frac{1}{4}-\xi^2)$ and  $m \to  \hat m/ {g_s} $.
Multiplying \eqref{dphi} by $g_s^2$ so that $\epsilon_1 =bg_s$ finite,
one has 
{\small
\be 
\left[ 
{\e_1^2}\left( \,z^2 \frac{\partial^2} {\partial z^2}
+z\frac{\partial}{\partial z} \right)
+\hat\Lambda^{2} \left(z^2+\frac1{z^2} \right)
+ \hat m \hat\Lambda\left(  z + \frac{1} {z} \right)
+\frac{\Lambda}{2} \frac{\partial}{\partial\Lambda} f_{\delta}\left(\hat\Lambda\right)
- \e_1^2\xi^2
\right]
\psi(z)   =0,
\ee}
where we use the limit $\lim_{g_s\to 0} g_s^2\frac{\Lambda}{2}\frac{\partial}{\partial\Lambda} \psi(\Lambda, z)  =0$. 

This equation can be considered on the unit circle  $z = {\rm e}^{2ix}$  with real $x$:
\begin{equation}
\label{preGM}
\left[-{\e_1^2}\frac{\textrm{d}^2}{\textrm{d}x^2}
+(8\hat\Lambda^{2}\cos4x +8\hat m \hat\Lambda\,\cos 2x)\right]\psi (x) = E \,\psi (x)\,,
\end{equation} 
where  $E=\;4\e_1^2\xi^2-2\hat\Lambda\,\partial_{\hat\Lambda}f_{\delta}\left(\hat\Lambda\right)
 $. 
This is the Schr\"odinger equation for $  \psi (\Lambda, x) $
with the potential real.
It is noted that we have the real potential since we put all the parameters of
$L$ and $R$  same: 
The $\psi$ corresponds to the expectation value of the irregular module.

\subsection{Differential equation and  loop equation} 
The same differential equation can also be  derived if one uses the  loop equation.
In fact, the derivation using the loop equation is simpler and can be easily generalized 
into higher rank cases.  
Let us define the conformal block with the degenerate operator 
$V_{-1/(2b) }(z)$ \cite{BMT_201104},
\be
    {\cal F}^{(n+m+2)}_{-1/(2b) }(z) 
  =   \left \langle\!\!\! \left\langle   V_{-1/(2b) }(z) \left( \int d \lambda e^{2b \phi(\lambda)} \right)^{\!\!N_+} 
       \prod_{0 \le k \le n+m+1}   V_{\a_k} (w_k)  
\right\rangle\!\!\!\right \rangle, 
    \ee
where $w_0 =0$ and $w_{n+m+1} \to \infty$. 
In addition,  $n $ number of operators  $w_k,  1\le k \le n $  are assumed  
to lie around $0$ and   $m $ number  of operators  
$w_k,  n+1\le k \le n+m $   around $\infty$
so that 
rank n (and m) colliding limit is obtained. 
Explicitly,
\ba
       {\cal F}^{(n+m+2)}_{-1/(2b) }(z)    
    &=&   
           \prod_{0 \leq k < \ell \leq n+m+1} (w_k - w_\ell)^{- \frac{2 \hat \a_k \hat \a_\ell}{g_s^2}} 
           \prod_{k=0}^{n+m+1} (z - w_k)^{ \frac{\hat \a_k}{b g_s}}
           \\
    & &    ~~~~~~\times
           \int \prod_{I=1}^{N_+} d \lambda_I \prod_{I<J} (\lambda_I - \lambda_J)^{- 2b^2}
           \prod_{I} \prod_{k=0}^{n-2} (\lambda_I - w_k)^{\frac{-2 b \hat \a_k}{g_s}}  (z - \lambda_I).\nonumber 
\ea
One may normalize the above with
$    {\cal F}^{(n+m+2)}
  =   \left \langle \!\!\!\left\langle
    \left( \int d \lambda e^{2b \phi(\lambda)} \right)^{\!\!N} 
       \prod_{0 \le k \le n+m+1}   V_{\a_k} (w_k)  
\right \rangle\!\!\! \right \rangle $.
However, one needs to care about the neutrality condition. 
For the expectation value one has the neutrality condition  
$ - 1/(2b) + \sum_k \a_k + N_+ b  =Q$ 
where as the partition function has the neutrality condition  
$   \sum_k \a_k + N  b  =Q$.
This requires that $N _+ - N = 1/(2b^2)$,
which shows that one needs different number of screening operators 
for the evaluation of the  partition function and for the expectation value. 
However, this unpleasant feature disappears when NS limit is achieved:
 $N _+ - N = 1/(2b^2) \to 0$ and one may identity $N _+ $ with $N$
and find the normalized degenerate conformal block as
\begin{align}
\frac{  {\cal F}^{(n+m+2)}_{-1/(2b) }(z)     }{  {\cal F}^{(n+m+2)}  }
&={\prod_{  k=0}^ { n+m+1}    (z - w_k)^{\hat\a_k/\e_1} }  \left\langle   \left(  \prod_I  ( z-\lambda_I ) \right) \right\rangle\,,
\end{align}
so that
 \ba
    \log  \left( \frac{  {\cal F}^{(n+m+2)}_{-1/(2b) }(z)     }{  {\cal F}^{(n+m+2)}  }\right)  
     =      \frac{V_{(m:n)}(z) }{\e_1} + \log \left< \prod_{ I} (z- \lambda_I) \right>.
 \ea
We introduce the bracket to denote the normalized expectation value 
of $\lambda_I$'s. \\
Define
$
 \eta (z) \equiv    \left\langle   \left(  \prod_I  ( z-\lambda_I ) \right) \right\rangle   \,,
$
noting that  \be   z - \lambda_I  =      C(\lambda_I;z_0) ~  e^ {   \int^{z}_{z_0} \frac{d z'}{z' - \lambda_I}} \,,
\ee
where $ C(\lambda_I;z_0)  $ is a $z$-independent normalization
one may put $ \eta(z)/ \eta (z_0)$  in terms of exponential form
{\it i.e.}, the irreducible effective action 
whose explicit form is given as 
\be
\log \left( \frac { \eta (z)}{ \eta (z_0)} \right) 
 =  
\sum_{k=1}^\infty \frac{1}{k!}  
\left\langle 
\left(  \sum_I\int_{z_0}^z  \frac {dz'} {z'-\lambda_I} \right)^k 
 \right \rangle_{\!\!\!conn}\,,
\ee
where the bracket with the subscript $c$ 
denotes the connected  part of the expectation value.  
This quantity is given in terms of the multi-point of the resolvent 
\be
\log \left( \frac { \eta (z)}{ \eta (z_0)} \right) 
 =   \frac1{  \beta } 
\sum_{k=1}^\infty \frac{1}{k!}   \left(  \frac{g}{\sqrt{\beta} } \right)^{k-2 } 
\int_{z_0}^z    \prod_{\ell=1} ^kdz'_\ell   ~
W(z'_1, z'_2, \cdots, z'_k)  \,,
\ee
where the multi-point resolvent is defined in \eqref{eq_multi-resolvent}.  
At the NS limit, all the multi-point resolvent vanishes except the one-point resolvent \cite{MMM}.
Therefore  the expectation value at the NS limit is given as 
\be 
\log \left( \frac { \eta (z)}{ \eta (z_0)} \right) 
 =  \frac1 {g \sqrt{\beta} }  
\int_{z_0}^z      dz' ~W(z' )  \,.
\ee
Recall that $x(z) = 2 W(z) + V'(z)$,  we find 
 \ba
    \frac{  {\cal F}^{(n+m+2)}_{-1/(2b) }(z)     }{  {\cal F}^{(n+m+2)}   }
     =      \Psi(z), ~~~
    \Psi (z)
     =     \exp \left( \frac{1}{\epsilon_1} \int^{z} x(z') dz' \right),
\ea
exactly the one we defined before.

Noting that the resolvent satisfies the loop equation 
we have the second order 
differential equation for $ { \Psi (z)}$  similar to the one 
in \eqref{schrodinger} 
  \be 
   \left ( \epsilon_1^2 \frac{\partial^2}{\partial z^2}   + U_{(m:n)}(z) \right)  \Psi (z) =0\,,
\ee 
with the potential $  U_{(m:n)} (z)$,
$NS$ limit of the potential $ V_{m:n}(z)$ 
\begin{align}
U_{(m:n)} (z)
 &=  - \Big (V_{(m:n)}'(z) \Big) ^2  - {\epsilon_1}  V_{(m:n)}''(z) - f(z) 
\nn\\  
&=    \sum_{k=-2m}^{2n} \frac {\tilde\Lambda_k}{z^{k+2}}
-\sum_{k=-m}^{n-1} \frac {\tilde d_k } {z^{2+k}} \,,
\end{align} 
where $\tilde \Lambda_k 
= (k+1) \epsilon_1 \hat  c_k-\sum _{\ell=-m}^{n} \hat c_\ell \hat c_{k-\ell}$ and according to \cite{CR},
\begin{align}
 -g_s^2~ v_k ( \log   {Z_{(m:n)}}) & =  \tilde d_k  \qquad\qquad~~~~~~~{\rm for}~0 \le k \le n-1\,,
\nn\\
 -g_s^2~ u_k ( \log   {Z_{(m:n)}}) & =  \tilde d_{-k}  -2 \epsilon_1 N \hat c_{-k} ~~~~{\rm for}~1 \le k <m-1\,,
\label{uv_k}
\\
2 \epsilon_1 N \hat c_{-m} & =  \tilde d_{-m}  \nn\,.
\end{align} 
Here $u_k$ is the differential operator corresponding to 
$\hat c_{-\ell}$,
$
 u_{k>0} =  \sum_{\ell >0}  \ell~\hat c_{-\ell -k }  \frac{ \partial}{\partial \hat c_{-\ell}} 
$.The potential has degree of poles  higher than 2 and non-vanishing zeros. 
Notice that the partition function $Z_{(m:n)}$ should also have the classical behavior $
Z_{(m:n)}
\;\stackrel{g_s \to 0}{\sim}\;
\exp\left\lbrace \frac{1}{g_s^2} \varsigma_{(m:n)} \right\rbrace.
$we have  Generalized Mathieu equation on the circle $z={\rm e}^{2ix}$ by
putting $  { \Psi (z)} =z^{-\Delta_+}    \psi ( z)$,
\begin{equation}
\label{GMathieu}
\left[-\frac{{\rm d}^2}{{\rm d}x^2}+
  \sum_{k=-2m}^{2n} \frac {4\t_k}{\e_1^2} \,{\rm e}^{-i2kx}
+
  \sum_{k=0}^{n-1} \frac {4v_k (\varsigma_{(m:n)})}{\e_1^2}{\rm e}^{-i2kx}
+
  \sum_{k=1}^{m-1} \frac {4u_k (\varsigma_{(m:n)})}{\e_1^2}{\rm e}^{i2kx} -1
\right]\psi(x)\;=\;0,
\end{equation}
where 
\ba
&& \tau_k=\left\{
\begin{array}{ll}
\tilde \Lambda_k \quad & 0 \le k \le  2n\\
\tilde \Lambda_k-2 \epsilon_1 N \hat c_{k}  \quad & -m \le k \le  -1\\
\tilde \Lambda_k \quad & -2m \leq k< -m\,.
\end{array}
\right.
\ea 
We see when $m=n$, with proper choice of $\t_k$, ${\rm e}^{-i2kx}+{\rm e}^{i2kx}$ will reproduce $2\cos2kx$.
\subsection{Example of the classical  irregular conformal block }  
We present here an explicit calculation of the classical conformal block for rank 1. 
Introducing new parameters which rescales the original quantities 
such  that  $E=\;\e_1^2\lambda$, $ h\;= {2\hat\Lambda}/\e_1$ and  $ M\;= {2\hat m}/\e_1$ 
we have \eqref{preGM}  as 
\begin{equation}
\label{GMathieu}
\frac{{\rm d}^2 \psi(x)}{{\rm d}x^2}+\left(\lambda - 2h^2 \cos 4x- 2hM \cos 2x\right)\psi(x) \;=\; 0,
\end{equation}
We are looking for  a quasi-periodic  solution $\psi(x)$  with a  Floquet exponent $\nu$  
\ba
\psi(x+\pi)\;=\;{\rm e}^{-i\pi\nu}\psi(x)\,,
\ea
where ${\rm e}^{-i\pi\nu}$ is called the Bloch factor.
We provide a brief procedure to solve the equation perturbatively 
 for small $h$ and $M$, 
whose method can be found  in \cite{MuellerKirsten_2012}. 
The solution may have value  $\lambda = \nu^2 -2h^2 \zeta $, 
where $\zeta$ is a small quantity. 
In this case we may rearrange the equation \eqref{GMathieu}  into the following form
\begin{equation}
\label{dv}
D_{\n}\psi=\left(2h^2\zeta + 2h^2 \cos 4x+ 2hM \cos 2x\right)\psi, 
\end{equation}
and use the perturbation in powers of $h$ and $M$.
Here $D_{\n}\equiv \frac{{\rm d}^2}{{\rm d}x^2}+\nu^2 $ is the ordinary differential operator
independent of $h$ and $M$.  

The lowest order  the solution $\psi^{(0)}\equiv \psi_{\n}$ has 
either of the form 
$ \cos \n x$, $ \sin \n x$, $e^{\pm i \n x}$ or their combinations.
 Using the product-to-sum formula (similar tricks also works for the exponential terms in \eqref{GMathieu}), we know $\psi_{\n}$ always satisfies that 
\begin{equation}
2 \cos( tx) \; \psi_{\n}=  \psi_{\n+t}+ \psi_{\n-t}.
\end{equation}
Inserting $\psi^{(0)}\equiv \psi_{\n}$ into
the right hand side of  \eqref{dv} we have 
\begin{equation}
R_{\n}^{(0)}=2h^2\zeta ~\psi_{\n}+ h^2( \psi_{\n+4}+\psi_{\n-4})+ hM ( \psi_{\n+2}+\psi_{\n-2})\,,
\end{equation}
which should be higher order than $h$. 
Thus we add the perturbation so that  $\psi = \psi^{(0)} + \psi^{(1)}$.
Since $D_{\n}\psi_{\n}=0$ and $D_{\n+t}\psi_{\n+t}=0$
we have 
$D_{\n}\psi_{\n+t}=-t(2\n+t)\psi_{\n+t}$
where we use the relation  $D_{\n+t}=D_{\n}+t(2\n+t)$.
Therefore, we can cancel the term 
 $\psi_{\n+t}$ in $R_{\n}^{(0)}$
by adding $\frac{\psi_{\n+t}}{-t(2\n+t)}$.
We may choose  
\begin{equation}
\psi^{(1)}= h^2\bigg( \frac{\psi_{\n+4}}{-4(2\n+4)}+\frac{\psi_{\n-4}}{4(2\n-4)}\bigg)+ hM \bigg( \frac{\psi_{\n+2}}{-2(2\n+2)}+\frac{\psi_{\n-2}}{2(2\n-2)}\bigg),
\end{equation}
which will cancel the last four terms in $R_{\n}^{(0)}$. 

Now we have  new contribution $R_{\n}^{(1)}$ in the right hand side of \eqref{dv} 
\begin{equation}
R_{\n}^{(1)}= h^2\bigg( \frac{R^{(0)}_{\n+4}}{-4(2\n+4)}+\frac{R^{(0)}_{\n-4}}{4(2\n-4)}\bigg)+ hM \bigg( \frac{R^{(0)}_{\n+2}}{-2(2\n+2)}+\frac{R^{(0)}_{\n-2}}{2(2\n-2)}\bigg)\,,
\end{equation}
which results in the combination (up to the terms proportional to $\psi_{\n}$):
\begin{equation}
R_{\n}^{(0)}+R_{\n}^{(1)}=\psi_{\n} \bigg\{ 2h^2\zeta +h^2M^2 \bigg( \frac{1}{-2(2\n+2)}+\frac{1}{2(2\n-2)}\bigg)+ {\cal O}(h)^4\bigg\} +\dots.
\end{equation}
This sum should be forced to vanish up to ${\cal O}(h^4)$, so we
have  $\zeta = - \frac{M^2}{4 \left(\nu ^2-1\right)}$.\\

Repeating the similar but lengthy calculation, 
we find the eigenvalue $\lambda$ as:
{\small
\begin{eqnarray}
\label{Floquet}
\lambda &=& \nu^2 +
h^2\frac{M^2}{2 \left(\nu ^2-1\right)}+
h^4\frac{\left(5 \nu ^2+7\right)M^4+24\left(\nu ^2-1\right)^2M^2+16\left(\nu ^2-1\right)^3}{32 \left(\nu ^2-4\right) \left(\nu ^2-1\right)^3}
+\dots.
\end{eqnarray}}
One may compare the definition of $\lambda$ with the eigenvalues obtained above  
order by order, using $f_{\delta}(\hat\Lambda) =\sum\limits_{n=1}\left(\hat\Lambda\right)^{\!\!2n}\!\!f_{\delta}^{(n)}$,
\begin{align}
\lambda
&= \nu^2 +
\frac{4{\hat\Lambda}^2}{\e_1^4}\frac{2\hat{m}^2}{ \left(\nu ^2-1\right)}+
\frac{16{\hat\Lambda}^4}{\e_1^8}\frac{\left(5 \nu ^2+7\right)\hat{m}^4+6{\e_1^2}\left(\nu ^2-1\right)^2\hat{m}^2+{\e_1^4}\left(\nu ^2-1\right)^3}{2 \left(\nu ^2-4\right) \left(\nu ^2-1\right)^3}
+\ldots\,.\nonumber
\\[10pt]
 & =4\xi^2 -\frac{2\hat\Lambda}{\e_1^2}\,\partial_{\hat\Lambda}
\left[\,\sum\limits_{n=1}\left(\hat\Lambda\right)^{\!\!2n}\!\!f_{\delta}^{(n)}\right]
\\[10pt]
&=4\xi^2-\frac{4{\hat\Lambda}^2}{\e_1^2}\,f_{\delta}^{(1)}-
\frac{8{\hat\Lambda}^4}{\e_1^2}\,f_{\delta}^{(2)}  - \ldots \nonumber
\end{align} 
To the lowest order one has $ \xi\;=\;\frac{\nu}{2}$. Hence $\d\equiv\e_1^2 (\frac{1}{4}-\xi^2)=\e_1^2(\frac{1}{4}-\frac{\nu^2}{4})$. 
We can read off the value of $f_{\delta}^{(n)}$ from the above, and they are indeed 
 consistent with those found in  \eqref{fdelta}.

It is observed that  if  we take  the limit $\hat m \to \infty$ and $h \to 0$ 
with $\hat m h = \tilde h^2$  constant, the above reduces to the
rank 1/2 ($N_f=0$ case) given in \cite{PP_201407}.

\section{Summary and discussion }  

In this paper we provide two ways to evaluate the classical limit of the irregular (2-point) conformal block 
with arbitrary rank. One is to take the direct limit from the irregular conformal block which is obtained 
using the irregular matrix model as presented in section 3. 
The classical irregular conformal block is given in an exponential form 
whose dominant contribution is checked by taking the classical limit 
of the irregular conformal block in section 3.2.

The other way is to solve the second order differential equation as given in section 4,
which is obtained by the null condition of degenerate primary operator. 
The differential equation is derived for arbitrary rank.
If one consider the expectation value of the degenerate primary field on the unit circle, 
then the equation turns out to be the generalized Mathieu equation
whose potential is given as the superposition of exponential (cosine) terms. 
We provide an explicit solution for the rank 1. 
The method is easily generalized for higher rank. 
It is noted that for rank $n \ge1$, there are $n$ number of coefficients 
which are given in terms of the differential form of the classical conformal block
with respect to the eigenvalues of $L_k$'s. 
This will provide the classical analogue of the flow equation obtained in \cite{CR}.

It is known  that the classical limit of the irregular conformal block 
is not simple to evaluate. 
However, the generalized Mathieu equation provides an alternative approach
to evaluate the classical conformal block in a systematic way. 
One may have  3-point conformal block with one degenerate primary field
still in terms of the generalized Mathieu equation whose potential is not real but  complex. 
In this case, it is more convenient to solve 
the differential equation on the complex plane 
rather than on a circle. The solution is given in Laurent series expansion of $z$ 
with a fractional power term attached. 
The series expansion can be done 
where potential terms are given as perturbation. 
It will be interesting to find the complete solution 
and to investigate its analytical structure. 
In this paper we only consider the case on the sphere but 
it is not hard to extend to higher genus case.
For the genus $1$, classical regular conformal block is discussed in \cite{BMT_201104}.

{\it Note added:} According to the referee report 
we add comments on a special limit 
of the irregular matrix model $Z_{(0;n)}$ 
which may  reduce to   $| {G_{m}} \rangle$ which appears in \cite{BMT_2011}.
Suppose one scales the parameters  $\hat c_k \to  q \hat c_k$ ($1 \le k \le n$) 
of the potential in \eqref {eq_potential-i-n}
and considers the limit $q \to 0$.
As a result all  $\Lambda_k \to 0$.
However,  the ratio  
$\hat c_\ell /\hat c_n =  c_\ell / c_n  $  $(1 \le \ell  \le n-1)$ is finite
and one may still  find the partition function  in terms of the ratios.
Note that $v_1$ and $v_{n-1}$ commute with each other $[v_1, v_{n-1}] =0$,
and one may regard $\tilde d_1$ and $\tilde d_{n-1}$ in \eqref{uv_k} 
as the eigenvalues of $v_1$ and $v_{n-1}$, respectively.
In this case, $\tilde d_1$ and $\tilde d_{n-1}$ are 
independent of  coefficients $\hat c_k$
and considered as  input  parameters 
where the filling fractions $N_i$ are given 
as functions of $\tilde d_1$ and $\tilde d_{n-1}$. 

In this framework, one may solve the equation  
\be 
v_k (-g_s^2 \ln Z_{0,n})=\tilde  d_k  ~~~{\rm for }~0 \le k \le n-1\,,
\label{eq_vk}
\ee
and $\tilde d_k$ (with $k \ne 1, n-1 $)  is to be found 
using the loop equation \eqref{eq_loop-i}. 
The solution will have the form
\be
-g_s^2 \ln Z_{(0;n)} 
= \tilde d_{n-1} h_{n-1}
+\tilde d_1 \left( \frac {c_{n-1}}{c_n (n-1)} \right)
 + g(\{c_k/c_n\})\,,
\ee
where $ g(\{c_k/c_n\})$ has to be determined by the condition 
$v_1(g) =v_{n-1} (g)  =0$ and $v_k (g)=\tilde d_k$ with $k \ne 1, n-1$
whose explicit solution is beyond the scope of this added note.  
On the other hand, $h_{n-1} $ 
has  explicit form which is the function of $c_k/c_n$'s and 
depends on $n$:
For example, when $n=2$, $ h_1$ reduces to $c_1/c_2$
and when  $n=4$,  
$  h_3 = \left( {c_1}/{c_4 } -  {c_2 c_3}/{(3c_4^2)}
+  {2c_3^3}/{(27c_4^3)}\right) $. 
In general when $n  \ge 3$, one has  
\be
h_{n-1} =  \sum_{k=1}^{n-2} \frac{c_k }{c_n }\bigg( \frac{ -c_{n-1}}{(n-1)c_n} \bigg)^{k-1}
-(n-2)\bigg( \frac{ -c_{n-1}}{(n-1)c_n} \bigg)^{n-1}\,.
\ee

It is also noted that 
the explicit form of irregular state $| {G_{m}} \rangle$
first constructed in \cite{BMT_2011}
is different from  $| {\widetilde G_{2n}} \rangle$ in  \eqref{G_2n}:
\be
| {G_{m}} \rangle =
\sum_{\ell,Y, \ell_p} 
\L^{2\ell/m}
\left\{ 
\prod_{i=1}^{[\frac{m}{2}]}  a_i ^{\ell_{2m-i} } 
\right\} \left\{ 
\prod_{j=1}^{[\frac{m-1}{2}]}   b_j^{\ell_j}  
\right\} 
Q_{\Delta} ^{-1} \Big(
 (m) ^{\ell_{m}}(m-1) ^{\ell_{m-1}}
\cdots 2^{\ell_2}1^{\ell_1} ; Y \Big)
L_{-Y} |\Delta \rangle \,.
\label{G_m}
\ee 
This $| {G_{m}} \rangle$ does not reduce to  $| {\widetilde G_{2n}} \rangle$ when $m=2n$.
The big difference is that the order of Young diagram in $Q_{\Delta} ^{-1}$
 is opposite to that of $| {\widetilde G_{2n}} \rangle$.
In fact $| {\widetilde G_{2n}} \rangle$ is the simultaneous eigenstate of $ {\cal L}_k$ 
with $n \le k \le 2n$. However, the  opposite ordering   makes 
$| {G_{m}} \rangle$ 
 simultaneous  eigenstate of ${\cal L}_1$ and ${\cal L}_{m}$:
 $  {\cal L}_1 \left| G_m \right>
     =     \Lambda^{\frac{2}{m}} b_{1} \left| G_m \right>$
and 
$   {\cal L}_m \left| G_m \right>
     =     \Lambda^2 \left| G_m \right>  $.  
This demonstrates that most of parameters except $\Lambda$ and $b_1$ 
given in \eqref{G_m} are  not fixed by the eigenvalue condition.  
The other parameters should be determined by solving the differential 
equation  \eqref{eq_vk}  as done in  \cite{CRZ} for  the case  $| \widetilde {G_{2n}} \rangle$.

\subsection*{Acknowledgements}
This work is supported by the National Research Foundation of Korea(NRF) grant funded by the Korea government(MSIP) (NRF-2014R1A2A2A01004951).

\end{document}